\documentclass[%
reprint, pdftex
amsmath,
amssymb,
aps,
pra
showpacs
]{revtex4-1}

\usepackage{graphicx}
\usepackage{dcolumn}
\usepackage{bm}
\usepackage[usenames,dvipsnames,svgnames,table]{xcolor}
\usepackage[normalem]{ulem}
\usepackage{upgreek}
\usepackage{physics}
\usepackage{hyperref}

\begin{document}
	
	\preprint{APS/123-QED}
	
	\title{Cavity-immune features in the spectra of superradiant crossover laser pulses}%
	
	\author{Mikkel Tang$^1$, Stefan A. Sch{\"a}ffer$^2$, Asbj{\o}rn A. J{\o}rgensen$^1$,  Martin R. Henriksen$^1$, Bjarke T. R. Christensen$^1$, J{\"o}rg H. M{\"u}ller$^1$, and Jan W. Thomsen$^1$}%
	\affiliation{$^1$Niels Bohr Institute, University of Copenhagen, Blegdamsvej 17, 2100 Copenhagen, Denmark\\
	$^2$Van der Waals-Zeeman Institute, University of Amsterdam, Science Park 904, 1098 XH Amsterdam, The Netherlands}

	\begin{abstract}
	    Lasing in the bad cavity regime has promising applications in precision metrology due to the reduced sensitivity to cavity noise. Here we investigate the spectral properties and phase behavior of pulsed lasing on the $^1$S$_0 - ^3$P$_1$ line of $^{88}$Sr in a mK thermal ensemble, as first described in \cite{qmetLasing}. The system operates in a regime where the Doppler-broadened atomic transition linewidth is several times larger than the cavity linewidth. We find that for some detunings of the cavity resonance, the influence of the cavity noise on the peak lasing frequency can be eliminated to first order despite the system not being deep in the bad cavity regime. Experimental results are compared to a model based on a Tavis-Cummings Hamiltonian, which enables us to investigate the interplay between different thermal velocity classes as the underlying mechanism for the reduction in cavity noise. These velocity-dependent dynamics can occur in pulsed lasing and during the turn-on behavior of lasers in the superradiant crossover regime.
	\end{abstract}
	
	\maketitle
	
	\section{Introduction}
	Lasers with stable, narrow frequency spectra have numerous applications in ultra-high precision measurements including state of the art clocks \cite{JILAlatticeClock, UshijimaLatticeClock, McGrewYbLatticeClock, SYRTEclocks, PTBclocks, INRIMclock, NPLclock} and gravitational wave detection \cite{LIGO, LISA}. These frequency references can be passive, in which case independently generated laser light is compared to an accurate reference. The state of the art performance of such systems is limited today by the thermal motion at the atomic level in the cavity mirrors \cite{Numata2004, Cole2013, MateiPRL2017, Robinson2019}. This limitation can be partially circumvented in an active reference, in which the system could consist of an atomic ensemble within an optical cavity \cite{KazakovEnsembles, KazakovColdAtoms, HollandLaser, ChenCaLaser, JingbiaoChen2005, ruggedLaser}. Here the idea is to pump the atoms to an excited state with a long lifetime relative to the decay of the cavity field, thus operating in the bad cavity regime \cite{ChenOE2019}. In this regime the cavity will enhance the emission rate by the atoms without pulling the frequency strongly to the cavity resonance, as the phase information is primarily stored in the atoms. This makes active frequency references operating in the bad cavity regime a promising technology for generating highly stable and reproducible frequencies, and has sparked interest in the physics of bad-cavity lasing.\\
	
	Superradiance can be observed when an atomic ensemble is coupled to a preferred electromagnetic mode. This process was originally investigated theoretically in the 1950s by R. H. Dicke \cite{Dicke}, and subsequently observed in the optical regime in 1973 \cite{HFsuperradiance}. Coupling an atomic ensemble to a broad cavity resonance, superradiant emission can be realized deep in the bad cavity regime. Superradiant pulses are characterized by a hyperbolic secant shape in the time-domain. The dynamics of superradiant lasing pulses have been studied in the context of narrow optical transitions on the 375 Hz wide $^1$S$_0 - ^3$P$_1$ line in Calcium \cite{CalciumDelayStatistics} and on the mHz wide $^1$S$_0 - ^3$P$_0$ transition in Strontium \cite{NorciaClockPulses}, where the effects of cavity pulling have also been characterized \cite{NorciaClockFreq}. In systems where the atomic transition and cavity linewidths are comparable, a superradiant crossover regime can be realized, characterized by both atomic coherence and a cavity field population, which acts back on the atoms. As a result these pulses deviate from the hyperbolic secant shape. The dynamics in this regime have been studied in experiments using the $^1$S$_0 - ^3$P$_1$ line in Strontium confined in an optical lattice at $\upmu$K temperatures \cite{NorciaCrossover} and in our system at mK temperatures \cite{qmetLasing}. Though lower temperatures imply less inhomogeneous broadening from Doppler shifts, these low temperatures are technically demanding to achieve. Thus a superradiant laser using hotter atoms could show great technological promise if the influence of Doppler shifts can be suppressed in the lasing frequency. Such a scheme has recently been proposed for operation in the continuous regime \cite{ruggedLaser} and reduces the technical challenges compared to continuous superradiance based on ultracold atoms \cite{KazakovColdAtoms}.\\
	
	In this paper we investigate the spectral properties and phase dynamics of an ensemble of unconfined $^{88}$Sr atoms emitting superradiant pulses of light via the mode of an optical cavity. In this system the temperature of the atoms results in a Doppler-broadened lasing transition linewidth that is larger than the cavity linewidth. Despite this, operating the system around a specific cavity detuning enables us to exploit the velocity-dependent behavior of moving atoms in order to suppress cavity pulling in the pulsed lasing regime. We present the experimental setup in Section II and the theoretical model of the system in Section III. This model is applied in the following sections to extract additional information about the physics at the atomic level, and simulations are shown along with the experimental results. We first describe the phase behavior of the lasing process in Section IV, then in Section V we present the spectral properties of the lasing pulses and relevant dynamics. In Section VI we analyze cavity pulling in the system and under which conditions this can be minimized, and in Section VII we evaluate the sensitivity of the system to cavity noise. In Section VIII we discuss improvements which are relevant for metrological applications.
	
	\section{Experimental system}
	
\begin{figure}[t]
		\includegraphics[width=\columnwidth]{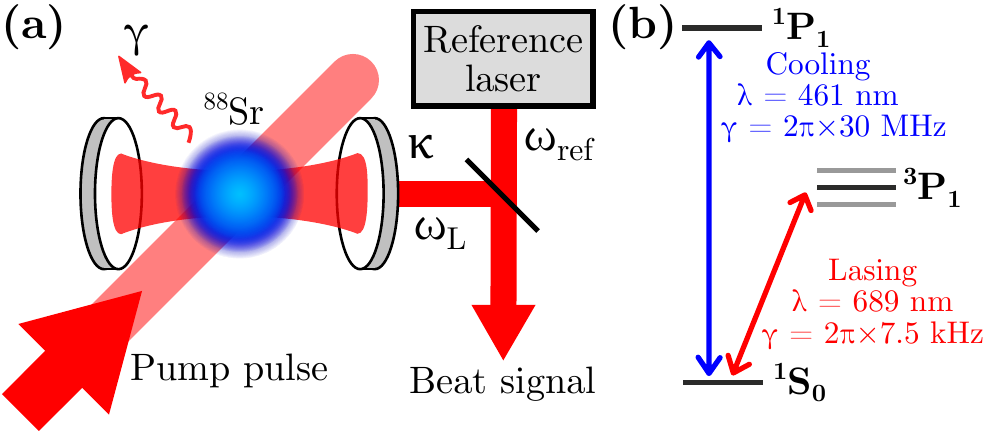}
		\caption{(a) Schematic of the experimental system showing a thermal ensemble of strontium atoms partially overlapping with the cavity mode. Coherent pumping prepares the atoms in $^3$P$_1$ and allows subsequent emission of a lasing pulse into the cavity mode. Light couples out of both ends of the cavity with the cavity linewidth $\kappa$. At one end the cavity output is overlapped with a reference laser beam, and the beat signal is detected. (b) Level scheme for $^{88}$Sr showing the cooling and lasing transitions. Wavelength ($\lambda$) and natural transition linewidths ($\gamma$) are indicated.\label{fig:setup}}
	\end{figure}

	The experimental system consists of a 5 mK ensemble of $^{88}$Sr atoms cooled and trapped in a 3D Magneto-Optical Trap (MOT) from a Zeeman-slowed atomic beam, using the $^1$S$_0 \leftrightarrow ^1$P$_1$ transition at 461 nm. The atomic cloud partially overlaps with an optical cavity and can be coherently pumped to $^3$P$_1$ using an off-axis pump laser, see Fig. \ref{fig:setup}. The cavity has a linewidth of $\kappa = 2\pi\times 620$~kHz and can be tuned to resonance with the $^1$S$_0 \leftrightarrow ^3$P$_1$ intercombination line ($\gamma = 2\pi\times 7.5$~kHz). For comparison the Doppler FWHM is $\Gamma_D = 2\pi \times 2.3$~MHz. Due to a cavity waist radius of $w_0 = 450$~$\upmu$m the number of atoms within the cavity, $N_{cav}$, is of order $10^7$, while the total number of atoms in the cloud, N, is of order $10^8$, as estimated from fluorescence measurements and lasing pulse intensity.\\
	
\begin{figure}[t!]
		\includegraphics[width=\columnwidth]{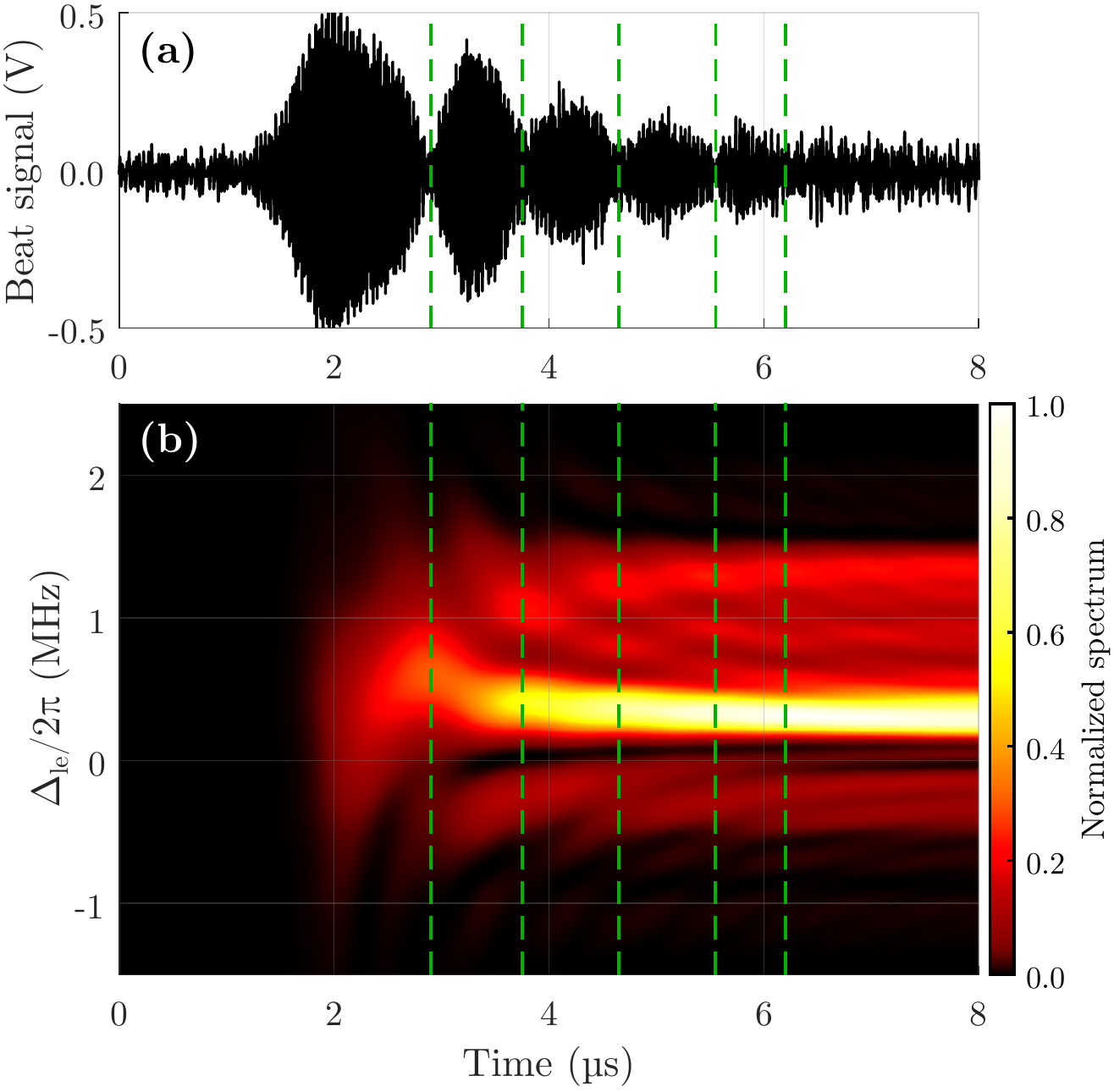}
		\caption{(a) Beat signal of a lasing pulse at a cavity detuning of 1 MHz. (b) Lasing intensity spectrum obtained by squaring the Fourier transformed beat signal for increasing window sizes, starting at $t = 0$. $\Delta_{le} = 0$ corresponds to the atomic transition frequency. Minima in emitted power are marked by green dashed lines.\label{fig:specEvolHeatmap}}
	\end{figure}

	Once the atomic cloud is established, the slowing and cooling beams are turned off, and the atoms are pumped to $^3$P$_1$ with an approximate $\pi$-pulse of 190 ns from an angle of 45$^\circ$ with respect to the cavity axis. Due to the large dimension of the cloud and the finite temperature, the population transferred to $^3$P$_1$ is limited to an estimated 85 \% within the cavity mode. Following a delay of a few microseconds, a lasing pulse is emitted by the atoms into the cavity. To characterize the spectrum of the laser pulses we beat the light leaking from one side of the cavity with a laser beam which is generated independently from the setup and stabilized to a transportable cavity on loan from SYRTE \cite{SYRTEcavity}. The reference laser frequency is detuned by 171 MHz from the atomic resonance, and the detected beat signal is then mixed at 120 MHz. This yields a signal with a frequency near 51 MHz which is band pass-filtered and recorded with a fast oscilloscope. Such a beat signal is shown in Fig. \ref{fig:specEvolHeatmap}a, for the case where the cavity is detuned by 1 MHz from the atomic resonance. This promotes oscillatory energy exchange between the atoms and cavity mode which results in damped oscillations in the emitted power. In Fig. \ref{fig:specEvolHeatmap}(b) a Fourier transformation is used to extract the spectrum for increasing window sizes, with the start fixed at $t = 0$. This method showcases how the spectrum changes throughout the lasing process; we see bifurcations each time additional intensity oscillations enter the analysis window. We also see that the spectral peak is shifted further towards the atomic resonance frequency as the window is expanded.\\
	
	\section{Modeling the laser spectrum}
	To model the laser pulses we numerically integrate a set of equations for the atomic and cavity states, derived in first order mean-field theory from a Tavis-Cummings Hamiltonian. The details of the model and lasing dynamics are described in \cite{qmetLasing}. To characterize the emitted spectrum in the model we extend the Hamiltonian with terms representing an array of imaginary filter cavities, following the method used by K. Debnath \emph{et al} \cite{Debnath}.
	
	\begin{eqnarray}
		H &=& \hbar \omega_c a^\dagger a + \sum_{j=1}^N   \hbar \omega_{e} \sigma_{ee}^j
		+ \sum_{k=1}^{N_f} \hbar \omega_{f}^{k} f_k^\dagger f_k
		\\\nonumber
		&+& \sum_{j=1}^N \hbar g_c^j  \left( \sigma_{ge}^j + \sigma_{eg}^j \right) \left( a + a^\dagger \right)\\\nonumber
		&+& \sum_{j=1}^N \hbar \frac{\chi_p^j}{2}  \left( \sigma_{ge}^j + \sigma_{eg}^j \right) \left( e^{ i \mathbf{k}_p \cdot \mathbf{r}_j - i \omega_p t } + e^{ -i \mathbf{k}_p \cdot \mathbf{r}_j + i \omega_p t } \right)\\\nonumber
		&+& \sum_{k=1}^{N_f} \hbar g_f  \left( a + a^\dagger \right) \left( f_k + f_k^\dagger \right).
	\end{eqnarray}
	
	Here the lowering operators $a$, $f_k$ and $\sigma_{ge}^j$ represent the cavity field, the $k$'th filter cavity, and the $j$'th atom, respectively. The angular frequencies are given by $\omega_{c}$ for the cavity, $\omega_{e}$ for the atom transition, $\omega_{p}$ for the pump pulse laser and $\omega_{f}^k$ for the $k$'th filter cavity. $\mathbf{k_p}$ is the pump pulse wavevector and $\mathbf{r_j}$ is the position vector of the $j$'th atom at a given time. $g_c^j$ is the position-dependent atom-cavity coupling, $\chi^j_p$ is the Rabi frequency of the semiclassical pump pulse laser, and $g_f$ is the coupling between the cavity and filter cavities.\\
	
	In this model we account for the motion and cavity coupling of each atom separately, due to the mK temperature and large spatial extent of the atomic cloud relative to the cavity waist. Since we describe expectation values our model neglects quantum noise in the system, which has little influence on the final spectrum due to the low Purcell decay rate [$\gamma_P = 1/($37 ms$)$] relative to coherent interactions. The filter cavities enable us to simulate the spectrum of the light leaking from the main cavity. Each of these have a very narrow linewidth $\kappa_F^j$ and is very weakly coupled (with coupling constant $g_f$) so that we can neglect any back-action on the lasing cavity. This appropriately models the experimental situation where laser photons do not re-enter the lasing cavity once they have left the output coupler towards the beat signal detector. The time evolution of the system operator expectation values then obey four distinct types of equations:
	
	\begin{eqnarray} \label{eq:OBE}
		\dot{\left\langle \sigma_{ge}^j \right\rangle} &=& -\left( i\Delta_{ep} + \frac{\gamma}{2} \right) \left\langle \sigma_{ge}^j \right\rangle \\\nonumber
		&&+ i \left( g_c^j \langle a \rangle + \frac { \chi_p^j}{2} e ^ { -i \mathbf{k}_p \cdot \mathbf{r}_j } \right) \left( \left\langle \sigma_{ee}^j \right\rangle - \left\langle \sigma_{gg}^j \right\rangle \right)\\\nonumber
		\dot{\left\langle \sigma_{ee}^j \right\rangle} &=&
		- \gamma \left\langle \sigma_{ee}^j \right\rangle + 
		i \left( g_c^j \left\langle a^\dagger \right\rangle + \frac{\chi_p^j}{2} e^{ i \mathbf{k}_p \cdot \mathbf{r}_j} \right) \left\langle \sigma_{ge}^j \right\rangle\\\nonumber
		&&- i \left( g_c^j \langle a \rangle + \frac{\chi_p^j}{2} e^{ -i \mathbf {k}_p \cdot \mathbf{r}_j } \right) \left\langle \sigma_{eg}^j \right\rangle.\\\nonumber
		\dot {\langle a \rangle} &=& 
		- \left( i \Delta_{cp} + \frac { \kappa } { 2 } \right) \langle a \rangle 
		- \sum_{j=1}^N i g_c^j \left\langle \sigma_{ge}^j \right\rangle\\\nonumber
    	\dot {\langle f_k \rangle} &=&
    	- \left( i \Delta^k_{fp} + \frac { \kappa^k_f } { 2 } \right) \langle f_k \rangle 
    	- i g_f \left\langle a \right\rangle.
	\end{eqnarray}
	
	Here the detunings $\Delta_{ij} = \omega_i - \omega_j$ have been introduced and a rotating frame has been chosen at $\omega_p$. The intensity spectrum from system initiation up to a time $t$ is then given by the relative populations of the filter cavities at time $t$ as a function of their resonance frequencies, $\abs{\expval{f_k(\omega_f^k)}}^2$, with a normalization depending on $g_f$. The number of filter cavities, $N_f$ and the range of frequencies they span determines the spectral resolution of the simulation.\\
	
	\section{Phase behavior}
\begin{figure}[t!]
		\includegraphics[width=\columnwidth]{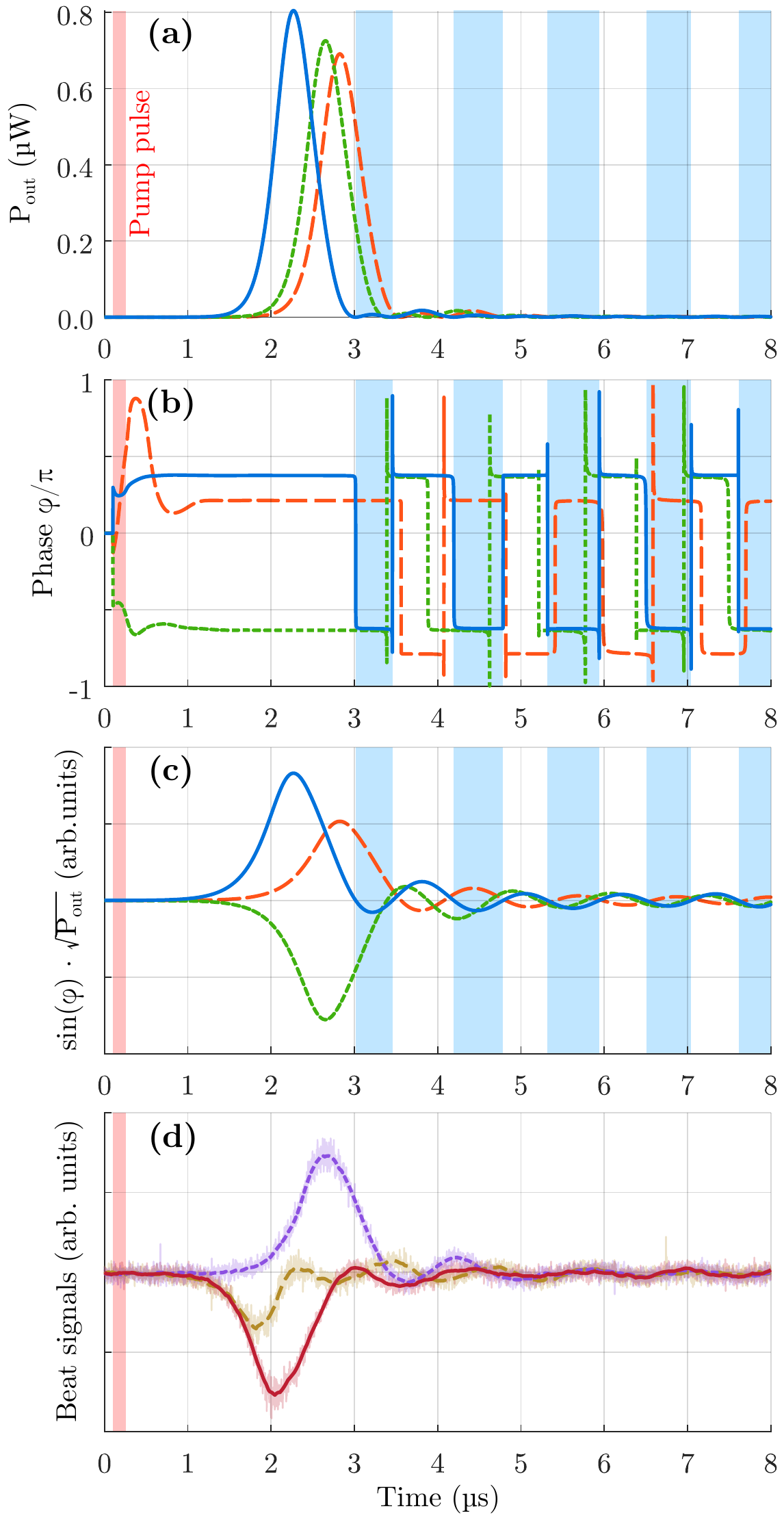}
		\caption{Time evolution of the cavity output power (a) and cavity field phase (b) for three simulated laser pulses after a pump pulse (on during the red shaded interval). The three instances have the same ensemble parameters but show random variation. (a) The laser pulses have random delays and are followed by small intensity oscillations when the cavity is on atom-resonance. The blue intervals mark every even number oscillation of the pulse with shortest delay. (b) The phase is random after the pump pulse, then evolves and stabilizes as the laser power builds up. For every intensity oscillation the phase flips sign. (c) Calculating $\sin{(\phi)} \cdot \sqrt{P_{out}}$ illustrates how a demodulated beat signal looks when measured in experiment. (d) Moving mean of three experimentally measured demodulated beat signals (shaded: raw data). We see qualitative agreement with the simulated behavior. \label{fig:detc0phaseEvol}}
	\end{figure}

    The phase coherence of the cavity field and its evolution is directly related to the spectral properties of the laser pulses and informs us about the underlying physical processes. The mean phase evolution can be traced in simulations from $\arg(\expval{a}$) in the semiclassical approximation. In Fig. \ref{fig:detc0phaseEvol} we show the simulated cavity output power for three random instances of lasing pulses when the cavity frequency is tuned to atom-resonance [Fig. \ref{fig:detc0phaseEvol}(a)] and the corresponding phase evolution of the cavity field [Fig. \ref{fig:detc0phaseEvol}(b)] in the rotating frame of the stationary atomic transition. The pump pulse initially imprints different phases on the Bloch vectors of all the atoms, depending on their positions and velocities. This prepares the ensemble with a non-zero macroscopic coherence, however it is tiny and randomized due to the 45$^{\circ}$ angle between pump pulse and cavity, and the random initial positions of the atoms at the wavelength level. Furthermore, in the semiclassical model a mean cavity field amplitude builds up during the pumping pulse corresponding to about 0.1 photons. Both the nonzero coherence and cavity field amplitude can separately lead to buildup of a lasing pulse. The atomic coherence can be removed after pumping experimentally by cycling the atoms on the $^1S_0$ - $^1P_1$ transition. If this is simulated, the remaining cavity field will imprint its random phase onto the atomic coherence as the power starts to build up. In the simulations we can also explicitly remove the field built up during pumping. In this case the randomized atomic coherence will trigger the dynamics. The resulting cavity field phase will depend on the macroscopically favored configuration originating from each atoms' state and coupling to the cavity mode. Keeping both the field built up during pumping and the initial random atomic coherence, the simulation shows both playing a role in the buildup of the laser pulse: after pumping, the phase is not stable, but the atoms build up coherence via the cavity field, which settles on a phase within 1 $\upmu$s after the pump pulse, as the lasing pulse starts to build up. At this point the output power has reached on the order of 0.1 nW, and the random delay between pumping and the lasing pulse is closely related to the time it takes for the phase to settle. With the choice of reference frame in Fig. \ref{fig:detc0phaseEvol}(b) a change in the phase of the cavity field over time corresponds to the cavity field being off resonance with the slowest atoms. This means the gain in the system is lower, slowing the pulse build-up and causing a longer delay of the pulse. A third process that can contribute to building up the lasing pulse is spontaneous emission into the cavity mode, which is not included in the model. With approximately $10^7$ atoms within the cavity waist, on the order of 60 photons are emitted into the cavity within 1 $\upmu$s given the Purcell rate of the system. This is the dominant process in triggering the lasing pulse in experiments, and is a source of phase noise that plays a role in the early stages of phase-synchronization.\\
    
    Though the ensemble preparation, scattering and spontaneous emission can initiate lasing and have different influences on the phase, the subsequent dynamics and spectral properties are highly deterministic. The Bloch vectors of the atoms in the ensemble all oscillate with different Rabi frequencies, depending on the atoms' velocities and positions with respect to the cavity waist and standing wave. Initially most atoms emit light, but at peak output power, most atoms start to reabsorb. By the end of the first intensity oscillation  [e.g. at $t = 3$ us for the blue traces in Fig. \ref{fig:detc0phaseEvol}(a) to (c)], the atoms are now out of phase due to their varying velocities and positions. The atoms that were absorbing at the end of the first lasing pulse begin to emit, and those that were emitting start to absorb, causing a $\pi$ phase shift of the second power oscillation. When the cavity is on atom-resonance, this pattern repeats during the subsequent intensity oscillations, as seen in Fig. \ref{fig:detc0phaseEvol}(b). We compare the simulations to measurements of the beat signal with an in-cavity reference laser on resonance with the cavity, but detuned 1 FSR from the atomic resonance, in Fig. \ref{fig:detc0phaseEvol}(c) (simulations) and \ref{fig:detc0phaseEvol}(d) (experiments). The simulation and experimental samples shown in this section are picked randomly from larger samples. We find that the measured beat signals also show the phase flipping behavior during the intensity oscillations, as the beat signals change sign between each oscillation.\\
	
\begin{figure}[t!]
		\includegraphics[width=\columnwidth]{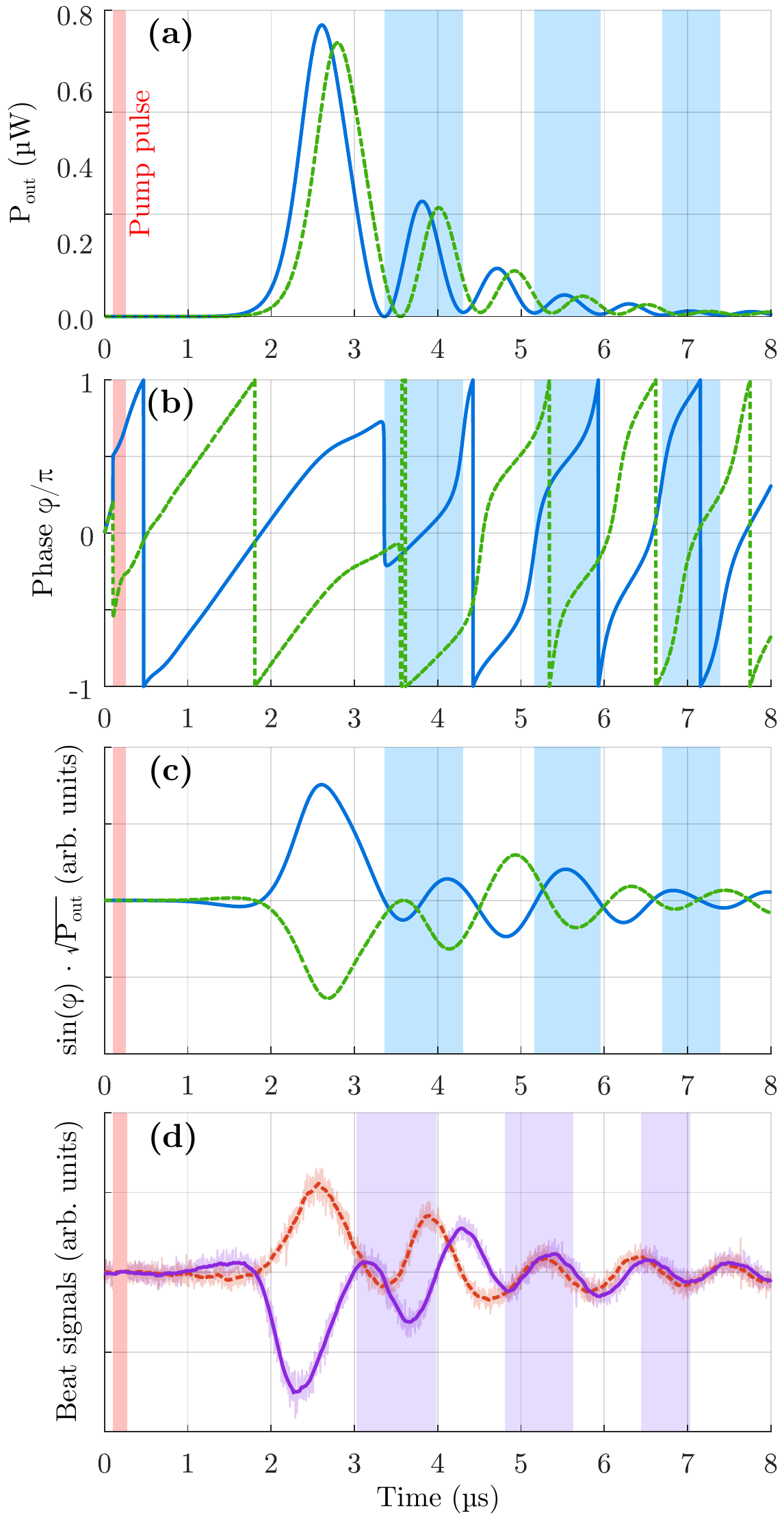}
		\caption{Time evolution of the cavity output power (a) and cavity field phase (b) for two simulated laser pulses after a pump pulse (on during the red shaded interval) at a cavity detuning of 1 MHz. (a) Oscillations in the output power are pronounced when the cavity is detuned. The blue areas mark every even number oscillation of the pulse with shortest delay. (b) The phase evolution during the laser pulses in a rotating reference frame at $\omega_c$. The sloped behavior is related to pulling of the lasing frequency away from the cavity resonance and towards the atomic transition frequency. (c) Calculated and (d) moving mean of experimental demodulated beat signals (shaded: raw data). Purple intervals mark every second oscillation in the experimentally measured output power. \label{fig:detc1MphaseEvol}}
	\end{figure}

	In the off-resonant case, where the cavity is detuned from the atomic resonance and pronounced after-pulsing is observed, the phase evolution is different (see Fig. \ref{fig:detc1MphaseEvol}). To compare the phase evolution with our experiments, the phase shown here has been transformed to a rotating frame at $\omega_c$. We find that the phase shift is close to $\pi$ between the first and second intensity oscillation, where the output power goes to zero. This explains the splitting of the spectrum in the experimental measurement of Fig. \ref{fig:specEvolHeatmap}(b) during the second oscillation, and the dip forming at the peak frequency of the initial pulse (these dynamics occur between the two leftmost green dashed lines). In subsequent oscillations the power does not go to zero and the phase shifts occur more gradually. The chirp in frequency towards the atomic resonance is also evident in the rate of change of the phase, which increases as the spectrum is pulled closer to atomic resonance.
	
	\section{Lasing spectra and underlying dynamics}
	
\begin{figure}[t!]
		\includegraphics[width=\columnwidth]{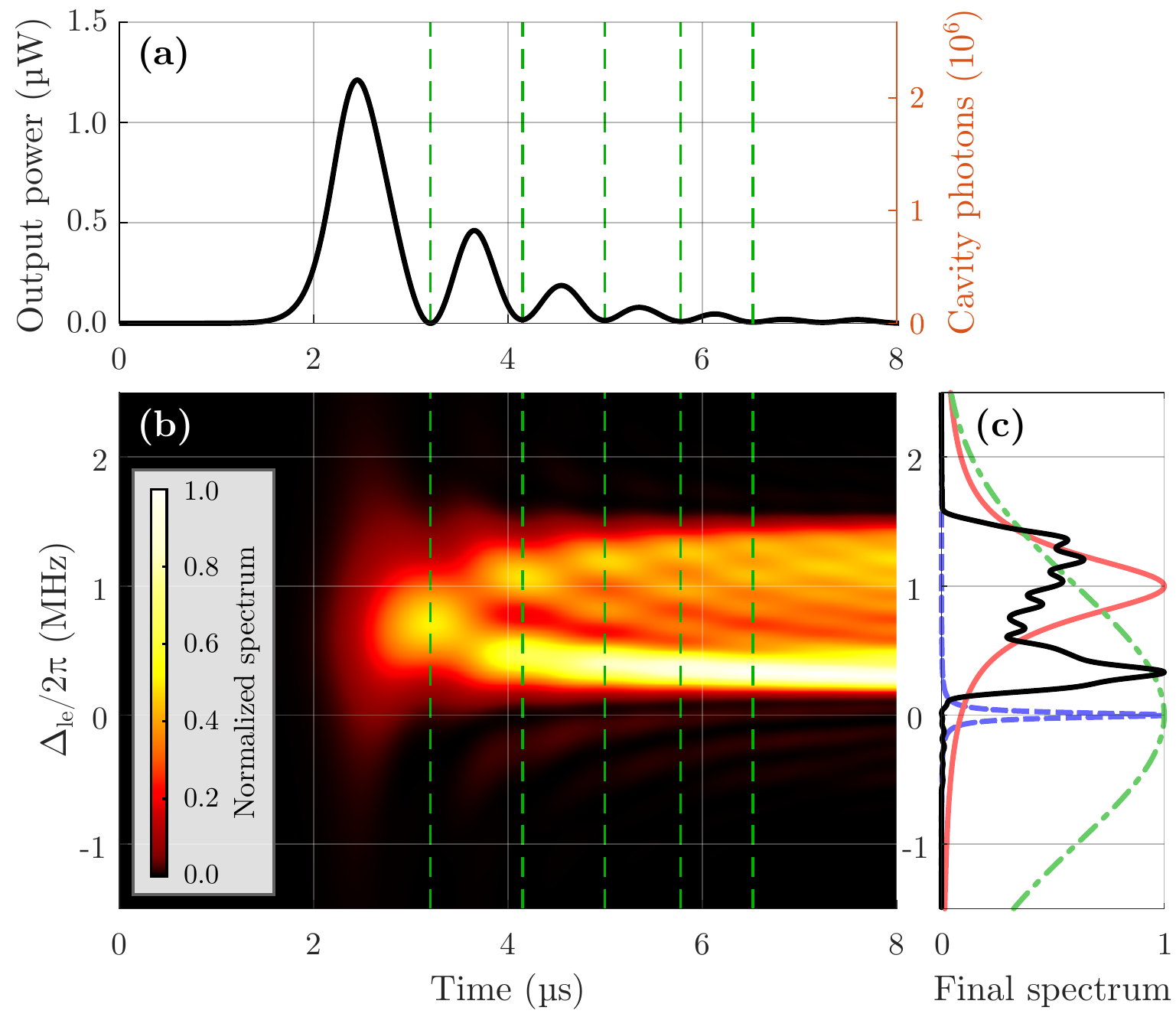}
		\caption{(a) Dynamics of the output power from one side of the cavity in a simulation, for a cavity detuning of 1 MHz. (b) Normalized lasing intensity spectrum as a function of window length in the simulation (starting at t = 0). As the intensity oscillations become included in the window, the peaks in the spectrum split up into multiple branches, and the dominant frequency is pulled towards the resonance frequency of the stationary atoms. Minima in output power are marked by green dashed lines. (c) The final spectrum at t = 8 $\upmu$s (black) compared to relevant lineshapes in the system (dashed blue: stationary atom, dash-dotted green: ensemble Doppler profile, red: cavity).\label{fig:specEvolHeatmapSim}}
	\end{figure}

	To compare simulations to our beat experiments in Fig. \ref{fig:specEvolHeatmap}, the simulated lasing dynamics of a pulse at a cavity detuning of 1 MHz is shown in Fig. \ref{fig:specEvolHeatmapSim}(a), and the corresponding evolution of the lasing spectrum in Fig. \ref{fig:specEvolHeatmapSim}(b). Here we recover the general structure found in Fig. \ref{fig:specEvolHeatmap}, including bifurcations and frequency chirp towards atom resonance. The pattern is found to closely resemble the velocity-dependent absorption/emission behavior previously reported in this regime \cite{qmetLasing} and is a fingerprint of the Doppler shifts of those atoms which interact strongly with the cavity mode. The final spectrum is shown in Fig. \ref{fig:specEvolHeatmapSim}(c), where the lasing spectrum is compared to relevant linewidths in our system: the natural linewidth $\gamma$ of a stationary atom, the Doppler linewidth of the ensemble, and the cavity linewidth shifted by $\Delta_{ce} = 2\pi \times 1$~MHz from the atomic resonance. Despite the large Doppler broadening and the cavity detuning of 1 MHz, the lasing spectrum has a peak component only 300 kHz away from the atomic resonance. This corresponds to the Doppler shift of atoms moving at 0.2 m/s along the cavity axis, which were previously shown to constantly emit during the intensity oscillations in this regime \cite{qmetLasing}. The width of the spectrum is Fourier limited, and thus highly dependent on the duration of the main laser pulse. As the atom number increases the pulse evolves faster, resulting in a shorter main laser pulse. Reducing the temperature of the atoms also leads to faster pulse evolution as it increases the atom number in the lower velocity classes. In addition the lasing peak near $\Delta_{le}$ = $2\pi \times 300$~kHz will become more prominent relative to the structures at higher frequencies. A less efficient pumping leads to a smaller pulse, similarly to the effect of having fewer atoms. Fluctuations in temperature, atom number and pumping efficiency will thus lead to variations in spectra in experiments, even if the cavity could be kept perfectly stable.\\
	
	\section{Cavity pulling}
	To characterize the sensitivity of our experimental system to cavity noise we measure the beat signals for a range of cavity detunings. We vary the cavity detuning in steps of 100 kHz in a randomly generated order to prevent systematic bias, and gather 100 beat signals for each cavity detuning. We find the frequency intensity spectra of individual laser pulses as shown in Fig. \ref{fig:specEvolHeatmap}.\\
	
	\begin{figure}[t!]
		\includegraphics[width=\columnwidth]{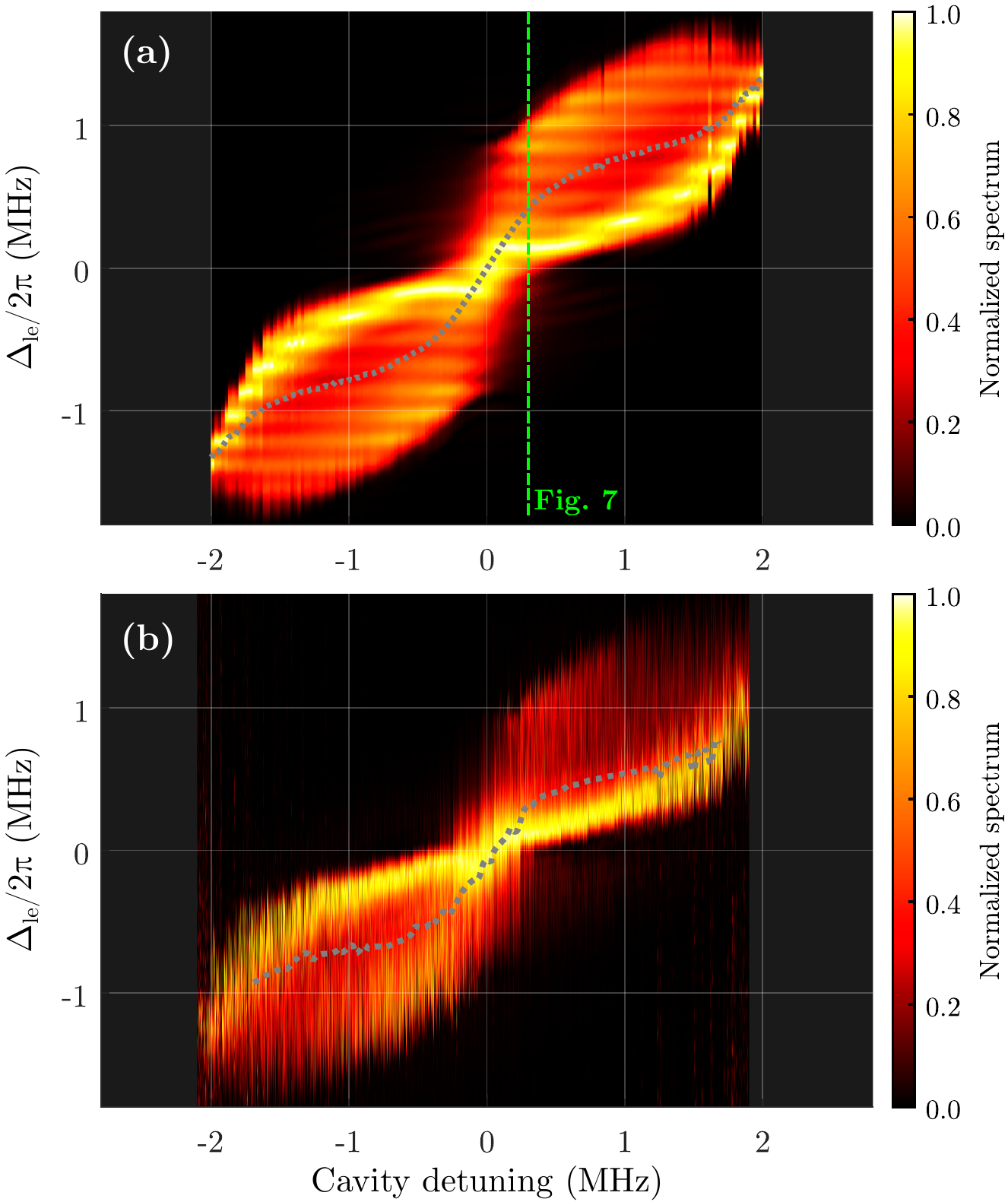}
		\caption{Intensity spectra from many lasing pulses for varying cavity detuning: (a) simulations, (b) experiments. Each column in the images constitute the spectrum of a single simulated/measured lasing pulse. The spectra are normalized such that the peak of each column equals one. The dotted gray lines indicate the center of mass of the lasing spectrum, which has a minimal slope of 0.25 at a cavity detuning of $\pm$1.2 MHz in panel a. The dashed green lines refer to the simulation in Fig. \ref{fig:velDatVertical} which illustrates the dynamics in this regime where location of the spectral peak is independent of cavity detuning. \label{fig:cavPull}}
	\end{figure}
	
	\begin{figure}[t!]
		\includegraphics[width=\columnwidth]{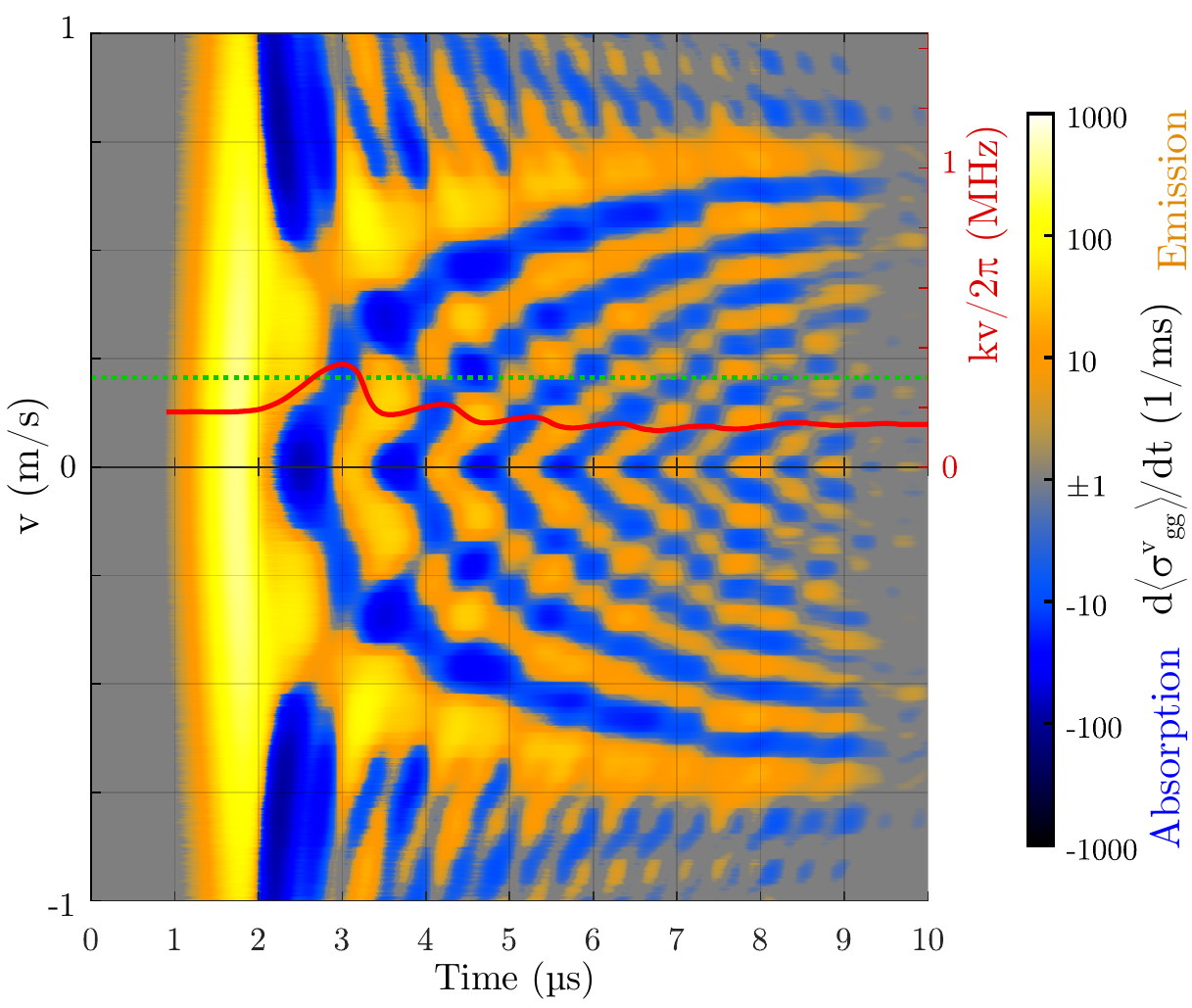}
		\caption{Simulated time evolution of the rate of emission (yellow) and absorption (blue) of cavity photons, depending on the velocity of atoms along the cavity axis. The dynamics are shown for a cavity detuning of 300 kHz. The dotted green line shows where the Doppler shift equals the cavity detuning on the right axis, and the red line shows the spectral peak frequency in comparison, with a window starting at t = 0.\label{fig:velDatVertical}}
	\end{figure}
	
	We can compare the experimentally measured spectra to simulated spectra, in order to understand the dependency on cavity detuning, see Fig. \ref{fig:cavPull}. We use a cavity detuning ($\Delta_{ce}/2\pi$) resolution of 25 kHz in the simulations [Fig. \ref{fig:cavPull}(a)]. In the experiments [Fig. \ref{fig:cavPull}(b)] 100 data samples are taken in steps of 100 kHz in nominal cavity detuning and distributed along the axis. Here the cavity detuning also fluctuates on the order of 100 kHz from its nominal value from shot to shot. We can characterize the immunity of the system to cavity noise by a local cavity pulling coefficient in terms of how much the lasing frequency changes when the cavity detuning is changed: $c_{pull} = d\Delta_{le}/d\Delta_{ce}$. The lasing spectrum can be partially characterized by its peak and center (defined by the center of mass) frequencies, which we will use to characterize cavity pulling. In an ideal active frequency reference the lasing frequency is completely immune to shifts in cavity resonance, which would yield $c_{pull}$ = 0, while in the good cavity regime where $\gamma \gg \kappa$, $c_{pull}$ approaches 1 as the cavity alone determines the lasing frequency. Simulations interestingly show that the peak frequency of the spectrum is immune to shifts in cavity resonance frequency for a range of detunings around $\pm$300 kHz, thus $c_{pull}^{peak}$ = 0. This local immunity to cavity fluctuations occurs because of the velocity-dependent interactions between the atoms and cavity in the pulsed lasing regime, as illustrated in the simulations in Fig. \ref{fig:velDatVertical}. The frequency of the initial lasing pulse starts out with a shift of about 2/3 of the cavity detuning, but during the subsequent cycles of reabsorption and reemission the spectral peak is pulled to 150 kHz off the atom resonance. In this interval the center frequency of the spectrum is still highly sensitive to the cavity detuning; $c_{pull}^{center}$ = 1. The minimal cavity pulling obtained when considering the center of the spectrum is $c_{pull}^{center}$ = 0.25 at $\Delta_{ce}/2\pi$ = $\pm$ 1.2 MHz, and in this range $c_{pull}^{peak}$ = 0.5. The dynamics in this regime are very similar to the behavior shown in Fig. \ref{fig:specEvolHeatmapSim}. Beyond detunings of 1.5 MHz cavity pulling begins to increase significantly as the cavity is so far away from resonance with the slowest atoms that their interaction becomes too weak for these atoms to filter the light. At these values the pulses start to become smaller, with fewer after-pulses than at 1 MHz, and the lasing process instead takes place mainly in a subset of fast-moving atoms.
	
	\section{Influence of cavity noise}
\begin{figure}[t!]
		\includegraphics[width=\columnwidth]{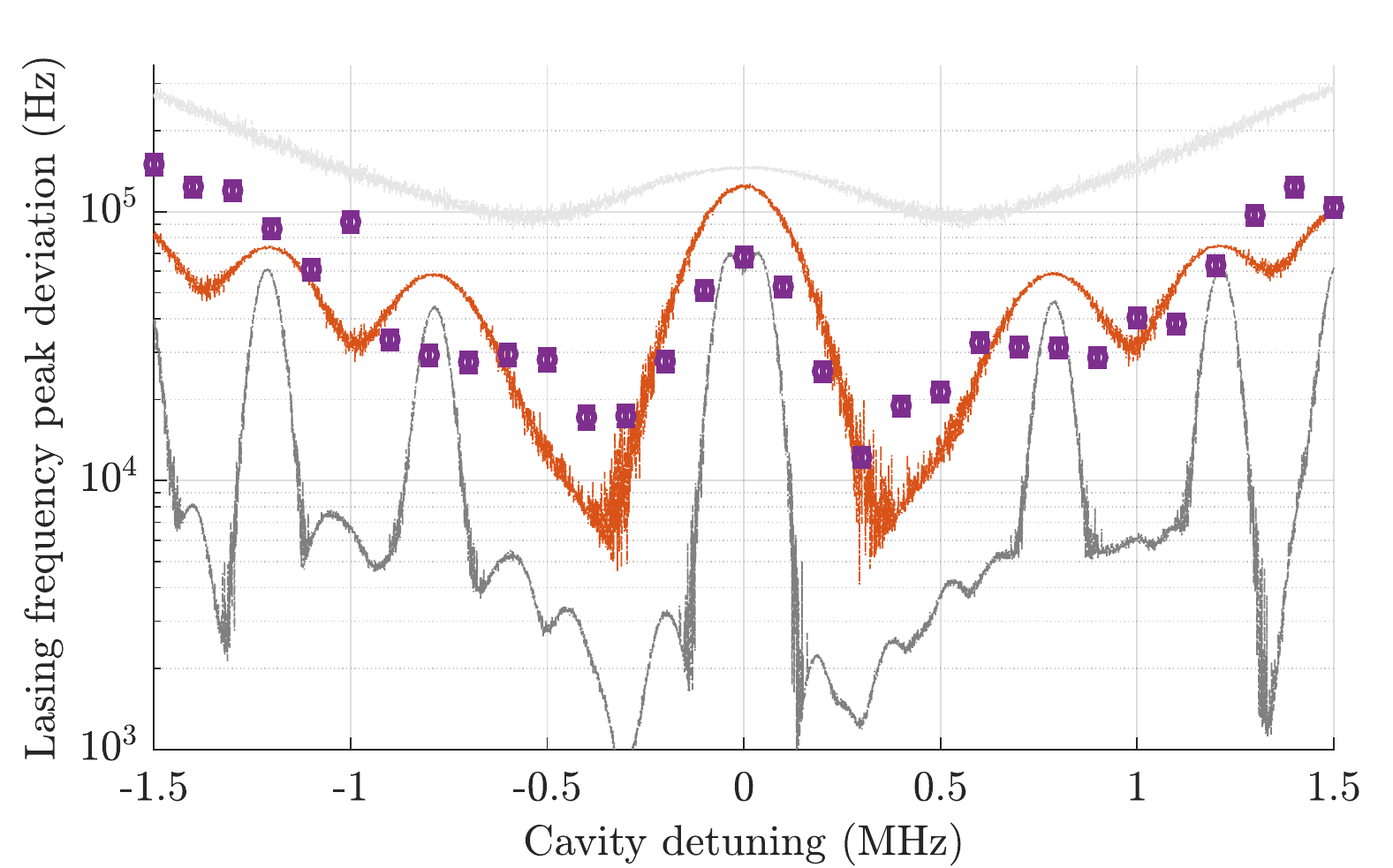}
		\caption{The sensitivity of the lasing peak frequencies found in experiments and simulations for different cavity detunings. Purple points show experimental frequency stability with errorbars indicating 1$\sigma$ uncertainty. The curves show simulations with added $\sigma_{jitter}$ of 30 kHz (dark grey) 100 kHz (orange) and 300 kHz (light grey). Variations in atom number and pumping efficiency causes additional variations in the peak frequencies found in experiments when compared to simulations. The fluctuations are minimal for a cavity detuning near $\pm$300 kHz. \label{fig:freqStab}}
	\end{figure}

	The non-linear dependency of cavity pulling on its frequency detuning suggests that the frequency stability of an active laser should also be dependent on the chosen cavity detuning during operation. This could allow for supression of classical noise from cavity fluctuations beyond the expected resonant cavity pulling factor. To investigate this quantitatively for the laser pulses shown in Fig. \ref{fig:cavPull} we extract the peak frequencies of the lasing spectra and plot their standard deviation in Fig. \ref{fig:freqStab}. We discard outliers based on a 5$\sigma$ criterium, which filters away data samples where e.g. the lasing pulse did not build up, and also some spectra for negative $\Delta_{ce}$ where the spectral peak occured in the bifurcating structure which is depicted in Fig. \ref{fig:specEvolHeatmapSim}(b) and (c) near $\Delta_{le}/2\pi$ = 1 MHz. Since the laser pulse often fails to build up for cavity detunings outside $\pm$1.5 MHz, the remaining analyzed samples in this regime will tend to be sampled from the lower range of the fluctuations, which are on the order of 100 kHz. To compare this to simulations we first assume that experimental fluctuations in the cavity detuning are the dominant contribution to the standard deviation of the experimental peak frequencies. This is supported by the fact that the frequency fluctuations are smallest near $\pm$300 kHz and largest near resonance or at large detunings, in agreement with the cavity pulling behavior. Therefore we interpolate the simulated peak frequency values as a function of cavity detuning, increasing the resolution by a factor of 50, and run a Monte Carlo-sampling of the interpolated peak frequencies obtained for a Gaussian distribution of cavity detunings with width $\sigma_{jitter}$. We pick 1000 points for each cavity detuning. The standard deviations of the obtained peak frequency distributions are shown in Fig. \ref{fig:freqStab} for $\sigma_{jitter}$ = 30, 100 and 300 kHz. We see that the experimental fluctuations are in reasonable agreement with the simulated ones for $\sigma_{jitter}$ = 100 kHz. For $\sigma_{jitter}$ = 30 kHz the lower values of the peak frequency deviations vary erratically. This is because the random variations in frequency from pulse to pulse start to become significant compared to the effect of cavity pulling.\\
	
	\section{Future improvements}
	Though the spectral peak frequency can be made highly insensitive to the cavity detuning, the spectral center of mass and other components still vary significantly with cavity detuning. By cooling the atoms to $\upmu$K temperatures (for example by additional cooling on the $^1$S$_0 - ^3$P$_1$ transition), the influence of Doppler shifts is reduced, and in this regime our simulations indicate that $c_{pull}^{peak}$ and $c_{pull}^{center}$ can simultaneously be reduced to 0.15, or separately reach 0 for different parameters. Here the initial inversion of the ensemble (which is determined by the pump pulse duration) also influences the lasing spectrum. This has previously been observed on the $^{87}$Sr clock transition in \cite{NorciaClockFreq}. This allows one to optimize both the cavity detuning and pump pulse duration to operate in a regime where the lasing frequency is approximately independent of the cavity frequency, and thus noise in the cavity.\\
	
	For metrological applications it is desireable to extend the emission time beyond a few microseconds. To reach the quasi-continuous or ultimately true continuous-wave lasing regime, where the gain is replenished while emission is taking place, incoherent repumping schemes could be employed, such as demonstrated in \cite{NorciaCrossover}. In these regimes the nature of the velocity-dependent dynamics and spectral properties would change significantly, as energy is replenished to emitting atoms with different velocities, or new atoms enter a populated cavity field. The velocity-dependent Rabi oscillations in the pulsed regime may on a longer timescale be replaced by synchronization among atoms within a range of velocities, similar to the dynamics simulated in \cite{DebnathFreqClasses}. The spectral properties due to these dynamics could have a very different dependency on the cavity detuning than the pulsed behavior and not necessarily lead to zero-crossings of the cavity pulling coefficient.
	
	\section{Conclusion}
	We have investigated the spectral and phase properties of pulsed lasing on the $^1$S$_0 - ^3$P$_1$ line in a mK Strontium ensemble both experimentally and in simulations, finding general agreement. We have described the processes involved in the random phase built up during the lasing pulses and investigated the phase-flipping behavior. In the resonant-cavity regime, the phase shifts abruptly by $\pi$ between each intensity oscillation, while the phase shifts are more gradual in the detuned-cavity regime. The spectral properties are complementary to the phase behavior, and we find that Doppler shifts play a significant role in the spectral properties. However even for a detuned cavity, slow atoms can pull the lasing frequency closer towards the stationary atomic resonance and suppress the influence of cavity drift on the spectrum. The extent of cavity pulling generally depends on a wide range of experimental parameters that can be optimized, including the cavity detuning and pump pulse duration. We notably find that the the peak spectral component is to first order insensitive to variations in cavity detuning for detunings around $\pm$300 kHz, due to the velocity-dependent absorption and emission dynamics in this regime. The presented simulation method, applied here to the pulsed regime of superradiance, can also be used to gain insight into turn-on and transient behavior of continuous superradiant lasers.
	
	\begin{acknowledgments}
	The authors would like to thank Rodolphe Le Targat and J\'er\^ome Lodewyck at SYRTE (Observatoire de Paris, CNRS, PSL, Sorbonne Universit\'e, LNE) allowing us to borrow a reference cavity and helpful advice in setting it up. They would also like to thank Yuan Zhang for useful theoretical discussions, and Sofus L. Kristensen, Valentin P. Cambier, Eliot A. Bohr and Julian C. R. Tait for useful discussions about the article. This project has received funding from the European Union’s (EU) Horizon 2020 research and innovation programme under grant agreement No 820404 (iqClock project), the USOQS projet (17FUN03) under the EMPIR initiative, and the Q-Clocks project under the European Comission's QuantERA initiative. We acknowledge support from VILLUM FONDEN via research grant 17558.
	\end{acknowledgments}

\end{document}